\documentclass[jap,graphicx, reprint]{revtex4-1} 
\usepackage{amsfonts, amsmath, graphicx}
\usepackage[version=3]{mhchem}

\begin{document} 
\title{Influence of Fermi arc states and double Weyl node on tunneling in a Dirac semimetal} 

\author{Zhuo Bin Siu}
\affiliation{Computational Nanoelectronics and Nanodevices Laboratory, Electrical and Computer Engineering Department, National University of Singapore, Singapore} 
\author{Can Yesilyurt}
\affiliation{Computational Nanoelectronics and Nanodevices Laboratory, Electrical and Computer Engineering Department, National University of Singapore, Singapore} 
\author{Mansoor B. A. Jalil} 
\affiliation{Computational Nanoelectronics and Nanodevices Laboratory, Electrical and Computer Engineering Department, National University of Singapore, Singapore} 
\author{Seng Ghee Tan} 
\affiliation{Data Storage Institute, Agency for Science, Technology and Research (A*STAR), Singapore} 

\begin{abstract}
Most theoretical studies of tunneling in Dirac and the closely related Weyl semimetals have modeled these materials as single Weyl nodes described by the three-dimensional Dirac equation $H = v_f \vec{p}\cdot\vec{\sigma}$. The influence of scattering between the different valleys centered around different Weyl nodes, and the Fermi arc states which connect these nodes are hence not evident from these studies. In this work we study the tunneling in a thin film system of the Dirac semimetal \ce{Na3Bi} consisting of a central segment with a gate potential, sandwiched between identical semi-infinite source and drain segments. The model Hamiltonian we use for \ce{Na3Bi} gives, for each spin, two Weyl nodes separated in $k$-space symmetrically about $k_z=0$. The presence of a top and bottom surface in the thin film geometry results in the appearance of Fermi arc states and energy subbands. We show that (for each spin) the presence of two Weyl nodes and the Fermi arc states result in enhanced transmission oscillations, and finite transmission even when the energy falls within the \textit{bulk} band gap in the central segment respectively.  These features are not evident in single Weyl node models. 
\end{abstract} 

\maketitle

\section{Introduction}

	The Dirac semimetal (DSM)  \cite{PRB85_195320, PRB83_205101, PRL108_140405, PRB88_125427} is a topologically non-trivial state which has attracted much attention recently.  Similar to the topologically protected surface states of the more well established three-dimensional topological insulators (TIs) \cite{RMP82_3045, JPSJ82_102001}, the low energy dispersion relations of DSMs take the form of Dirac cones. Differing from the 3D TIs where there are odd numbers of Dirac cones within the Brillouin zone and the linear dispersion of the Dirac fermion Hamiltonian hold only in two dimensions for the surface states,  the Weyl nodes in the \textit{bulk} states of DSMs disperse linearly in three dimensions. and appear in pairs. One Weyl node of each pair is a source of Berry curvature while the other node is a Berry curvature sink. Fermi arcs linking the two members of each pair emerge when a bulk DSM is truncated and a surface introduced perpendicular to the $k$ space separation between the Weyl nodes introduced. These Weyl nodes are topologically stable against perturbations which preserve the translational symmetry. To date, two materials, \ce{Cd3As2} \cite{NatMat13_677, PRL113_027603, NatComm5_3786, NatMat13_851} and \ce{Na3Bi} \cite{Sci343_864, APL105_031901, Sci347_294}  have been experimentally confirmed to host the DSM state. 
	
	Most of the existing studies on tunneling in DSMs and the closely related Weyl semimetals WSMs have focused on infinitely sized bulk DSM / WSM slabs in the the $k$ space vicinity around a single Weyl node \cite{APL108_163105,SciRep6_21283,PLA380_764}. This neglects the effect of inter-valley tunneling between multiple Weyl nodes. The influence of Fermi arc states on the tunneling process  is also not evident in these models as the explicit appearance of Fermi arcs requires pairs of Weyl nodes separated in $k$-space to be considered, as well as the presence of surfaces perpendicular to the $k$ space separation between the arcs. 
	
	In order to elucidate how multiple Weyl nodes and Fermi arc state on the tunneling spectrum compared with single Weyl node models of DSM/WSM tunneling, we study in this work the tunneling through thin films of the DSM \ce{Na3Bi} \cite{PRB91_121101,PRB94_235127,SciRep5_7898,Ar1601_05538} in an experimental setup similar to that in the earlier works, most of which have focused on Klein tunneling. We study the transmission from a semi-infinitely long source DSM thin film segment to a semi-infinite drain DSM thin film segment through a central segment subjected to a potential. 
	
	We mode the \ce{Na3Bi} using the Hamiltonian of Refs. \cite{PRB85_195320,PRB88_125427}
\begin{equation}
	H = \epsilon_0(\vec{k}) + \begin{pmatrix} M(\vec{k}) & A k_+ & 0 & 0 \\
	Ak_- & -M(\vec{k}) & 0 & 0 \\
	0 & 0 & M(\vec{k}) & -Ak_- \\
	0 & 0 & -A k_+ & - M(\vec{k}) \end{pmatrix} \label{H0}
\end{equation}
where  $\epsilon_0 = C_0 + C_1 \pi_z^2 + C_2 \pi^2$ and $M(\vec{\pi}) = M_0 - M_1 \pi_z^2 - M_2 \pi^2$ \cite{PRB85_195320, PRB88_125427}, $\pi^2 = \pi_x^2+\pi_y^2$ and $\pi_\pm = \pi_x \pm i \pi_y$.  $A$, $C_0, C_1, C_2, M_0, M_1$ and $M_2$ are material parameters for which we used the values in Ref. \cite{PRB85_195320}. The Hamiltonian consists of two uncoupled blocks representing the spin up and spin down states. Since the spin up and spin down states are uncoupled we can study each of these spin states separately. Here we focus on the spin up states. (The corresponding results for the spin down states can be obtained from those for the spin up states by replacing $k_y \rightarrow -k_y$. ) 

	This Hamiltonian yields, for each spin and an infinite-sized homogeneous \ce{Na3Bi} crystal, two Weyl nodes located at $k_x=k_y=0$ an at $k_x = \pm\delta k_z$. In the infinite crystal the dispersion in the vicinity of each Weyl node is linear,  reminiscent of a Dirac cone and can be identified with a `valley'. 
	
\begin{figure}[ht!]
\centering
\includegraphics[scale=0.5]{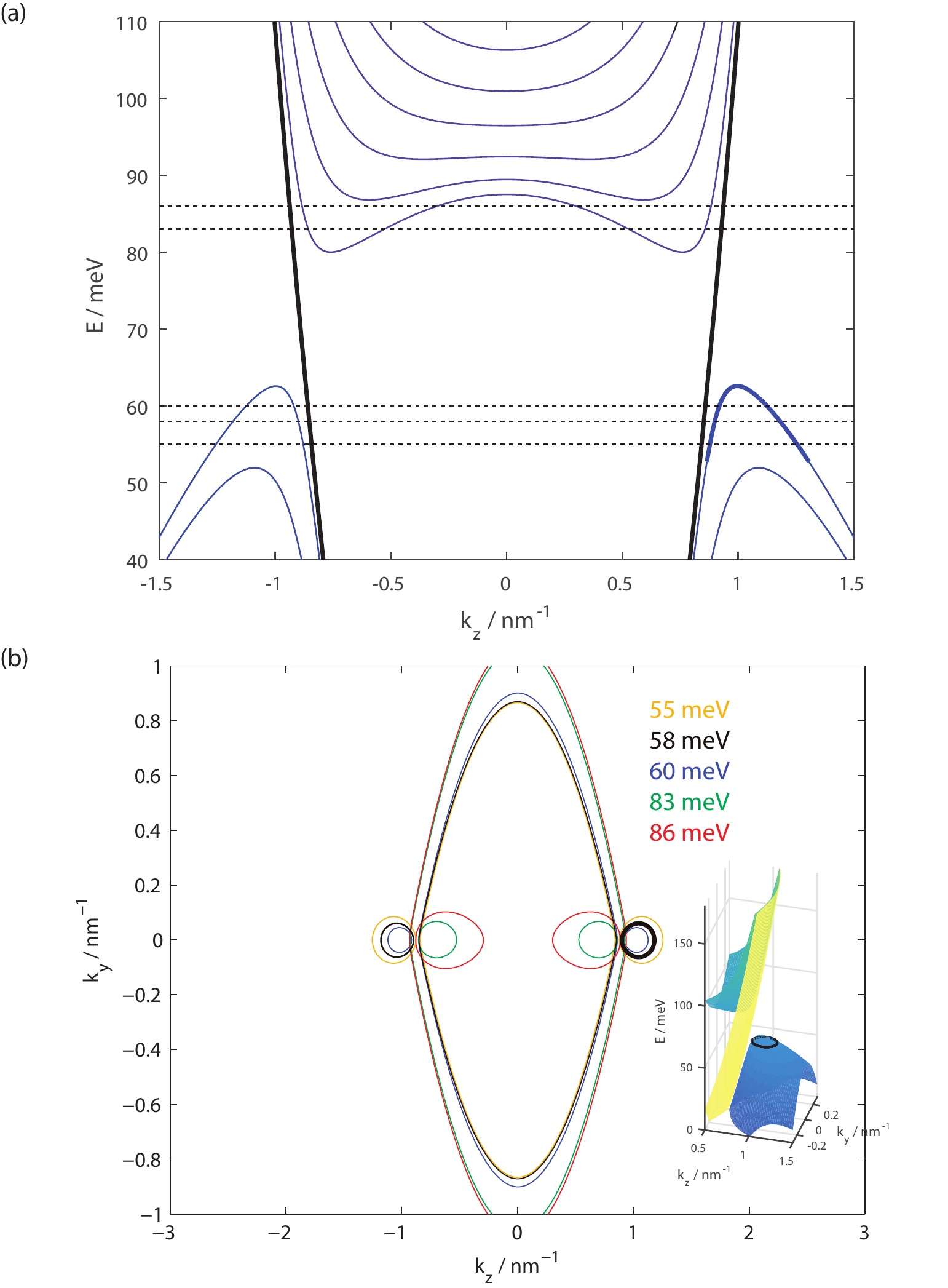}
\caption{ (a) The dispersion relations at $k_y = 0$ for a $50\ \mathrm{nm}$ thick \ce{Na3Bi} thin film. The bands in black correspond to the Fermi arc states. (b) The equal energy contours (EECs) for the same thin film at various values of energy indicated in the figure legend, and in panel (a) by the dotted lines at the corresponding values of energy. The inset shows the three dimensional plot of the gapped Dirac  with positive $k_z$ with the black ring around the hole like states of the cone depicting the source states whose transmission we will study in this paper. The yellow sheet between the upper and lower cones correspond to the Fermi arc states.    } 
\label{gW50Paper1}
\end{figure}

	The introduction of the thin film geometry with infinite dimensions along the $y$ and $z$ directions and finite thickness $W$ along the $x$ direction lead to the formation of energy subbands  (Fig. \ref{gW50Paper1}a ) due to the quantum confinement along the $x$ direction. (The eigenspectrum was obtained by expanding the  Schroedinger equation $H|\psi\rangle = |\psi\rangle E$ in the basis of the $\langle x|\psi\rangle = \sqrt{\frac{2}{W}} \cos(n\pi x/W)$ infinite potential well eigenstates and numerically diagonalizing the resulting matrix. )    
	
	The subbands, except for the specific bands indicated in black in panel (a) of the figure, are bulk subbands in the sense that there is significant particle density in the interior of the thin film away from the top and bottom surfaces. The cross sections of these bulk states on the $E-k_z$ dispersion graphs of panel (a) correspond to the elliptical cross sections of the Dirac cones at small $|k_y|$ distributed symmetrically about $k_z=0$ in panel (b).  The subbands leads to the opening up of a bulk energy gap between the top of the hole bulk bands and the bottom of the particle bulk bands.  The presence of the top and bottom surfaces at $x=\pm W/2$ results in the appearance of Fermi arc states \cite{PRB91_121101,PRB94_235127}, indicated in black in panel (a) of the figure, as well as in panel (b). These Fermi arc states differ from the bulk bands in that they are localized near the top or bottom surface of the thin film -- for a given spin the arcs convex in the $+k_y$  ($-k_y$) direction are localized near the top (bottom) surface, and exist even inside the bulk energy gap.  
	
	We note a few features in the EECs of Fig. \ref{gW50Paper1} that we will need later to explain the transmission profile.  The figure corresponds to a \ce{Na3Bi} thin film of thickness $50\ \mathrm{nm}$. This will be the thickness of the thin films we study in the remainder of this paper.  The finite thickness of the film leads to the opening of a bulk energy gap between  $63\ \mathrm{meV}$ to $79\ \mathrm{meV}$ crossed only by the Fermi arc states. For the particular model Hamiltonian and parameter set of \ce{N32Bi} which we are employing, the elliptical cross sections of the Dirac cones at the two valleys shift inwards towards smaller values of $|k_z|$ as the energy increases. 
	In order to connect with earlier single Weyl node studies of tunneling in WSM / DSM materials we consider the energy and gate potential ranges when there is only ta single propagating bulk band per valley and / or the Fermi arc state in the source and drain leads, and the central segment, and consider transmission from only the $+$ve $k_z$ valley bulk states. (These states are highlighted in the black ring in the inset of panel (b) of Fig. \ref{gW50Paper1}b. ) We set the lead energy at $58\ \mathrm{meV}$ as this gives a relatively wide range of gate potential in the central segment where only a single bulk band and / or the Fermi arc states are propagating.  We consider cases where the interfaces between the central segment and the source/drain leads are parallel, as well as perpendicular to the $k$ space separation between the Weyl nodes. We shall see that the presence of the two valleys as well as the Fermi arc states lead to the emergence of features missed in the earlier studies. 
	
\section{Transmission}
\begin{figure}[ht!]
\centering
\includegraphics[scale=0.5]{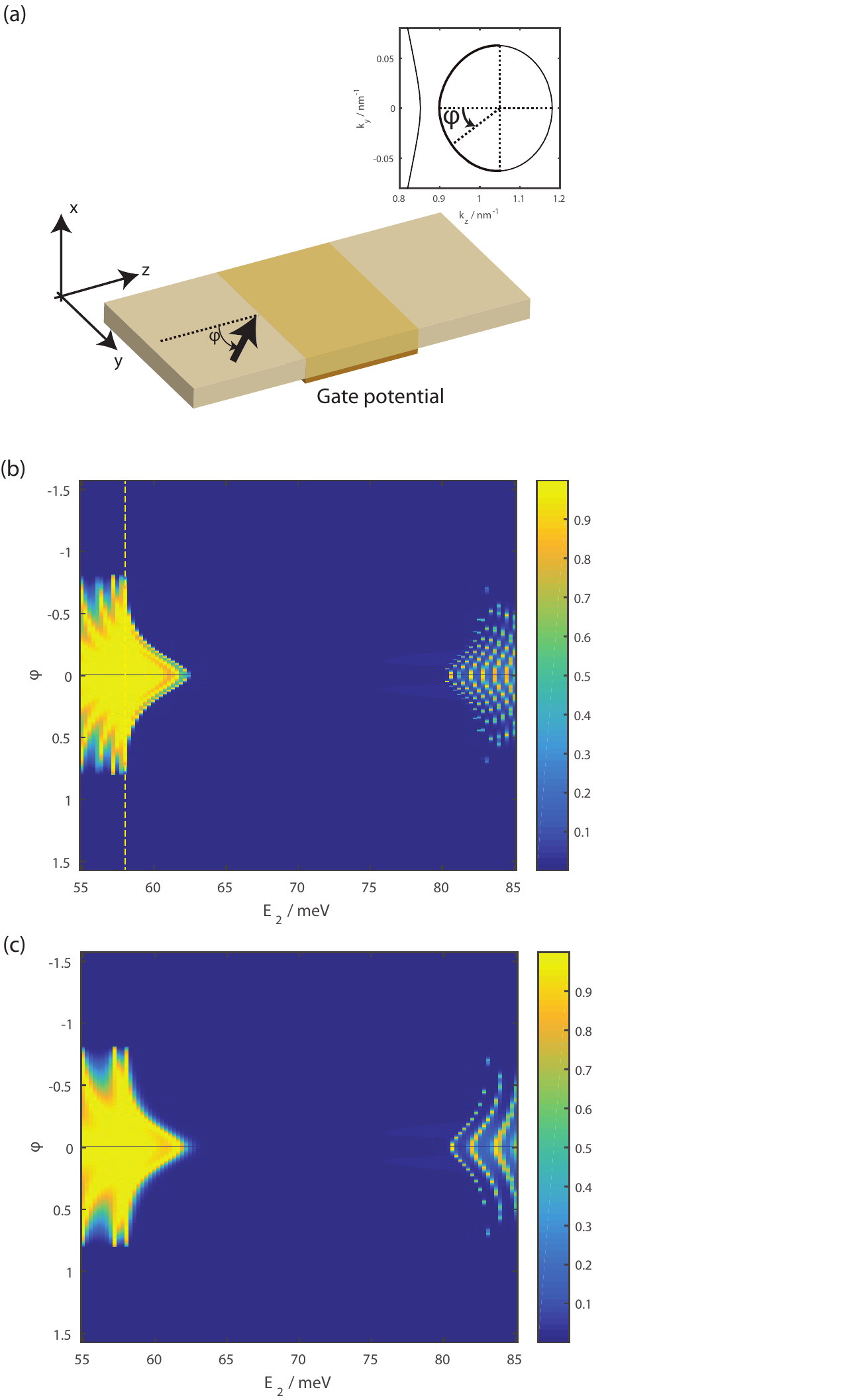}
\caption{ (a)  A schematic of the tunneling setup consisting of a central segment subjected to a gate potential sandwiched between semi-infinite long source and drain leads. (b) and (c) show the transmission as a function of the angular coordinate $\varphi$ along the positive $k_z$ Dirac `cone'  (see inset of panel (a) ) for definition of $\varphi$, and $E_2$ in the central segment for a central segment of length (b) 100 nm and (c) 50 nm.  $\varphi$ is roughly equal to the incident angle of the source current incident on the source-central segment interface as indicated in the main diagram of panel (a). The vertical yellow dotted line in panel (b) indicates the energy of the source and drain segments.   } 
\label{gTunKySchem}
\end{figure}

	We first study the transmission from the positive $k_z$ valley hole-like states at $E_1 = 58\ \mathrm{meV}$ source segment, through the central segment to the drain segment when the interfaces between the source and central segment, and between the central segment and the drain, are parallel to the $y$ axis.  The system is hence translationally invariant along the $y$ direction and $k_y$ is conserved.   The gate applies a potential which shifts the effective energy of the central segment $E_2 = E_1+ U$ with respect to that of the source and drain leads. Fig. \ref{gTunKySchem} shows the transmission as a function of $E_2$ and the angular coordinate about the source Dirac cone $\varphi$ for two different lengths of the central segment. 
	
For $E_2 < E = 58\ \mathrm{meV}$, the transmission is relatively high with a moderate amount of transmission oscillation for $-\pi/2 < \varphi < \pi/2$. As $E_2$ increases between $58\ \mathrm{meV}$ to $63\ \mathrm{meV}$ the $\varphi$ angular region where there is relatively high transmission tapers down to a point. This range of $E_2$ corresponds to the transmission from the hole states in the source to the hole states in the central segment. The transmission is almost (but not exactly) 0 for all values of $\varphi$ for $E_2$ ranging from $63\ \mathrm{meV}$ to $75\ \mathrm{meV}$. Between $75\ \mathrm{meV}$ to $80\  \mathrm{meV}$ there are two small angular ranges on either side of $\varphi = 0$ where there is small but significant transmission. Finally as $E_2$ increases above $80\ \mathrm{meV}$ the transmission becomes large again, with the $\varphi$ angular range over which the transmission is large increasing with $E_2$ until it spans the range from $-\pi/2 < \varphi < \pi/2$. The transmission for $E_2 > 80\ \mathrm{meV}$ corresponds to that from the hole states in the source to the particle states in the central segment. The transmission oscillations for $E_2 > 80\ \mathrm{meV}$ are more prominent and occur with a shorter $E_2$ periodicity compared to the fringes at $E_2 < 63\ \mathrm{meV}$. These transmission oscillations in turn have a shorter $E_2$ periodicity for the longer segment compared to the shorter one.

\begin{figure}[ht!]
\centering
\includegraphics[scale=0.3]{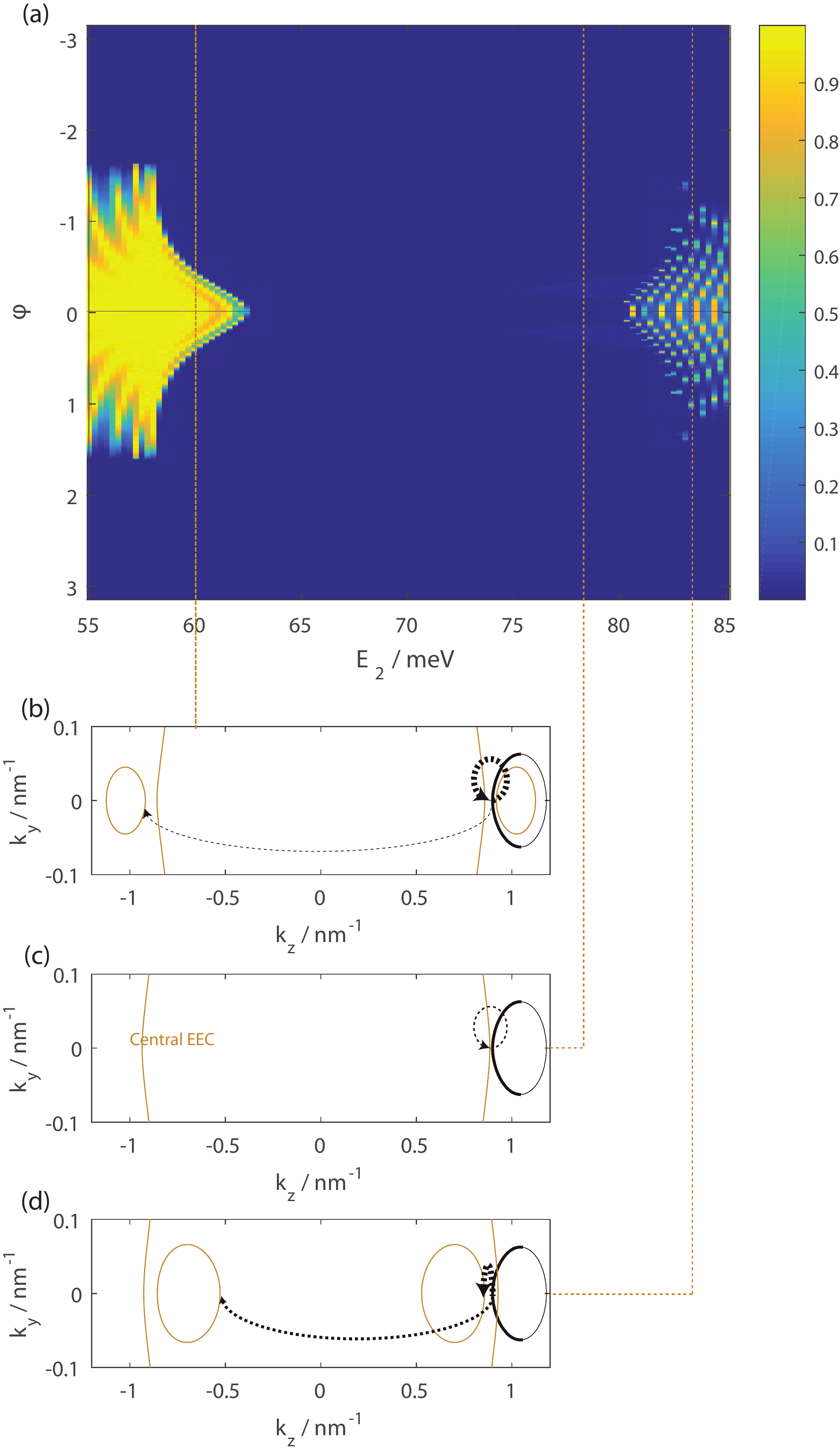}
\caption{ Panel (a) is the same plot of the transmission against angular coordinate as in Fig. \ref{htunKySchem}(a). Panels (b) to (d) profile show the source (thick black lines) and central segment (thinner green lines) EECs at the indicated values of $E_2$ in panel (a). The relative thickness of the dotted arrows pointing from points on the source EEC to the central segment EEC at $k_y=0$ indicate the relative probabilities of the $k_y=0$ central segment states propagating in the $+z$ direction the source $k_y=0$ state gets transmitted into. The thicker sections of the source EECs correspond to the source states which progagate in the $+z$ direction.  } 
\label{gW50tunY}
\end{figure}

These features may be explained by $k_y$ conservation and the EEC profiles in the leads (i.e. the source and drain), and the central segments as shown in Fig. \ref{gW50tunY}. 

$k_y$ conservation restricts the angular range over which there is finite transmission to the range of $\varphi$ in which there are propagating states for the values of source $k_y$ which $\varphi$ corresponds to. In general, the transmission probability from a bulk source state to a central segment Fermi arc state, while finite, is relatively very low compared to the transmission probability to a central segment bulk state. This is largely due to the lack of spatial overlap between wavefunctions of the Fermi arc states, which are localized either surface of the film, and the bulk states where significant probability density is found within the interior of the film.  For $E_2 < 58 \mathrm{meV}$, the elliptical cross sections of the bulk `Dirac cones' span a larger $k_y$ range than the $k_y$ range spanned by the bulk source EEC. The angular range of $\varphi$ with high transmission is thus limited by the $k_y$ range spanned by the source EEC and the  transmission is relatively high over the entire range of $-\pi/2 < \varphi < \pi/2$ over which the source states propagate in the $+z$ direction. Between $58\ \mathrm{meV} < E_2 < 63\ \mathrm{meV}$ the $k_y$ range spanned by the elliptical cross sections of the central segment Dirac hole cones shrinks with increasing $E_2$ to 0 at the bottom of the bulk energy gap. This leads to the narrowing of the range of $\varphi$ with relatively high transmission to zero. We note from panel (b) that for $E_2 < \mathrm{meV}$ there is a small probability for a source state from the positive $k_z$ valley to be transmitted to a central segment state at the negative $k_z$ valley.  (This probability is still significantly higher than that for transmission to a Fermi arc state. ) This inter-valley transmission contributes to the oscillation of the transmission at fixed $\phi$ as the energy varies.

The $E_2$ range between $63\ \mathrm{meV}$ to $80\ \mathrm{meV}$ corresponds to the bulk gap in the central segment where the only propagating states are the Fermi arc states. The transmission of the bulk source states to the the central Fermi arc states give a small but finite transmission over this $E_2$ range that becomes especially prominent from $75\ \mathrm{meV}$ to $80\ \mathrm{meV}$. 

As $E_2$ increases beyond $80\ \mathrm{meV}$ the $\varphi$ range with high transmission increase with energy as the $k_y$ range spanned by the central segment EECs increases to 0 to beyond the $k_y$ range spanned by the source states. Compared to the $E_2 < 63\ \mathrm{meV}$ energy range, there is an increased probability of the positive $k_z$ source states being transmitted to the other valley with negative $k_z$. This may be due to the closer pseudospin alignment between the hole states in the positive $k_z$ valley on the source side, and the particle states in the negative $k_z$ valley on the central segment side, than the pseudospin alignment between the hole states in different valleys when $E_2 < 63\ \mathrm{meV}$. This increased inter-valley transmission results in more prominent oscillations of the transmission at a fixed $\varphi$ as the energy is varied beyond $E_2 > 80 \ \mathrm{meV}$ compared to the oscillations at $E_2 < 63\ \mathrm{meV}$.   

The inter-valley scattering and the Fermi arcs are absent in single Weyl node models.  The enhanced Fabry-Perot transmission oscillations, as well as the small angular ranges with small finite transmission occurring at energy ranges falling within the central segment bulk gap due to inter-valley scattering and Fermi arc states respectively will hence not be captured in these models.

\begin{figure}[ht!]
\centering
\includegraphics[scale=0.6]{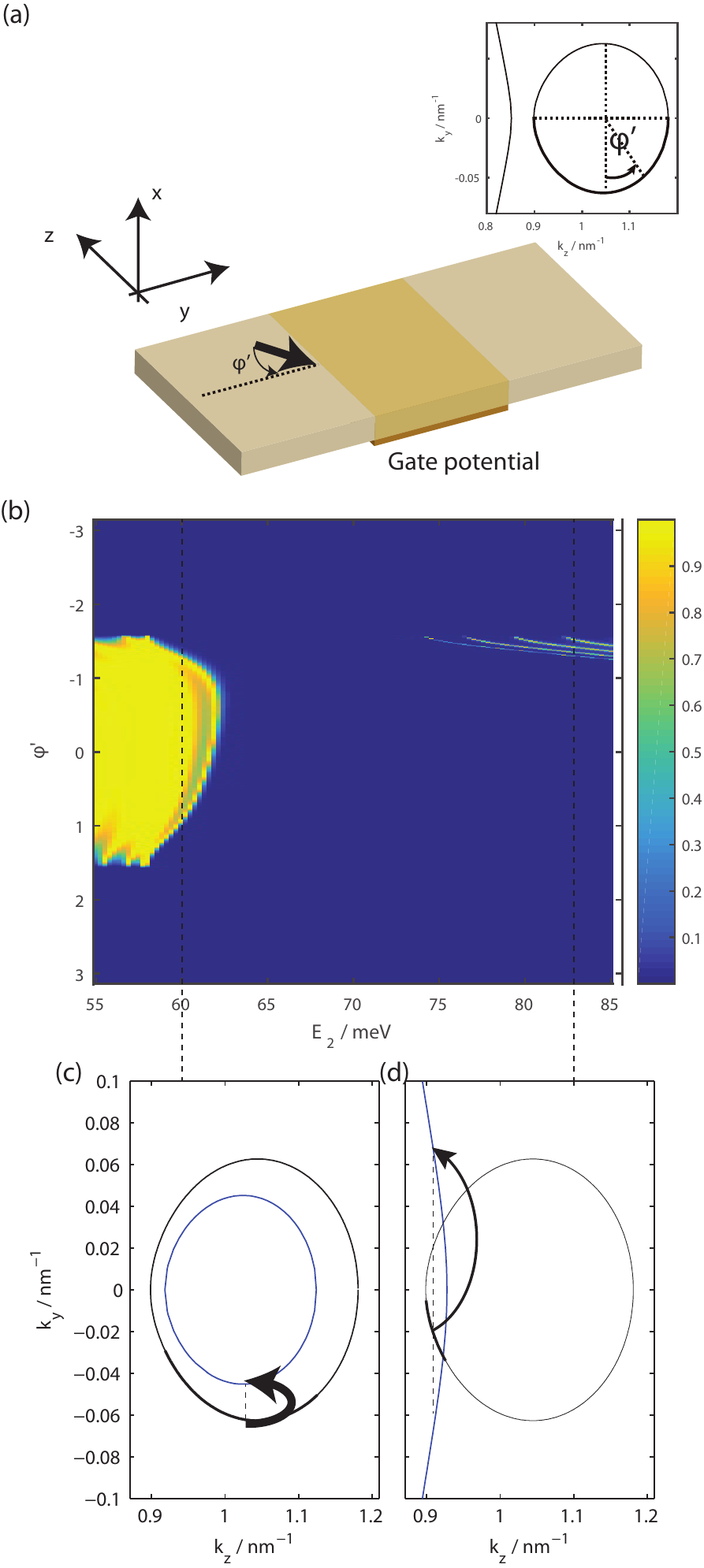}
\caption{ (a) The transmission as a function of the source angular coordinate $\varphi'$, and the central segment energy $E_2$.  $\varphi'$ is defined in the inset of the panel where the states propagating in the $+y$ direction are indicated on the thicker line. (b) and (c) show the source (black) and central segment (blue) EECs at the $E_2$ values indicated on the plot. The states on the source EEC propagating in the $+y$ direction with ranges of $k_z$ that overlap with the $k_z$ range of states on the central segment propagating in the $+y$ direction are indicated as thicker lines. The thickness of the arrows pointing from the source to the central segment EECs in (b) and (c)  show the relative probability of transmission form the source states to the central segment states at the value of $k_z$ indicated by the dotted lines.  } 
\label{gCA1comb}
\end{figure}

As a comparative example, we now turn our attention to the transmission from the source to the drain segment through a central segment subjected to a gate potential when the central segment-lead interfaces are now along the $y$ direction. In this case, $k_z$ is conserved. The transmission profile is somewhat more difficult compare directly against single Weyl-model models as the centers of the elliptical cross sections of the `Dirac cones' here shift inwards towards smaller $|k_z|$ values with increasing energy (Fig.\ref{gW50Paper1} ). This inwards shift leads to distinct features in the transmission profile, shown in Fig. \ref{gCA1comb}. (Note that $\varphi'$ is defined differently from $\varphi$ in the earlier figures in order for $\varphi'=0$ to correspond to normal incidence of the source wavefunctions on the source-central segment interfaces. ) 

Unlike the case where the lead-central segment interfaces are parallel to the $y$ direction, the transmission for interfaces parallel to the $z$ direction are not symmetrical about $\varphi'=0$. This asymmetry is due to the $k_z$ shift of the center of the elliptical Dirac cone cross sections towards $|k_z|=0$ with increasing energy. This shift results in the central segment state with the same pseudospin direction as the source state at a given value of $k_z$ being the central segment state with a slightly displaced value of $k_z$. Another major difference is that there is 0 transmission to the other valley here since the $k_z$ ranges of the two valleys do not overlap. The lack of inter-valley transmission results in less prominent transmission oscillations with energy variation compared. 

Similar to the interface along $z$ case studied earlier, the transmission is relatively high for $-\pi/2 < \varphi' < \pi/2$ for $E_2 < 58\ \mathrm{meV}$. The angular range of $\varphi'$ with high transmission  drops to 0 as $E_2$ is increased beyond $58\ \mathrm{meV}$. This drop can be attributed to the combined effects of the shrinkage and displacement of the central EEC segment with increasing energy, both of which reduce the range of $k_z$ overlap between the source and central segment EECs. Between $64\ \mathrm{meV} < E_2 < 75\ \mathrm{meV}$ the transmission is zero as there are no propagating central segment states within the $k_z$ range spanned by the source EEC.   As $E_2$ increases beyond $75\ \mathrm{meV}$ the central segment Fermi arc starts to move into the $k_z$ span of the source EEC, giving finite transmission again. The relatively large $k_y$ difference between the central segment Fermi arc state and source bulk state at a given value of $k_z$ results in more pronounced transmission oscillations with energy variation.  

\begin{figure}[ht!]
\centering
\includegraphics[scale=0.6]{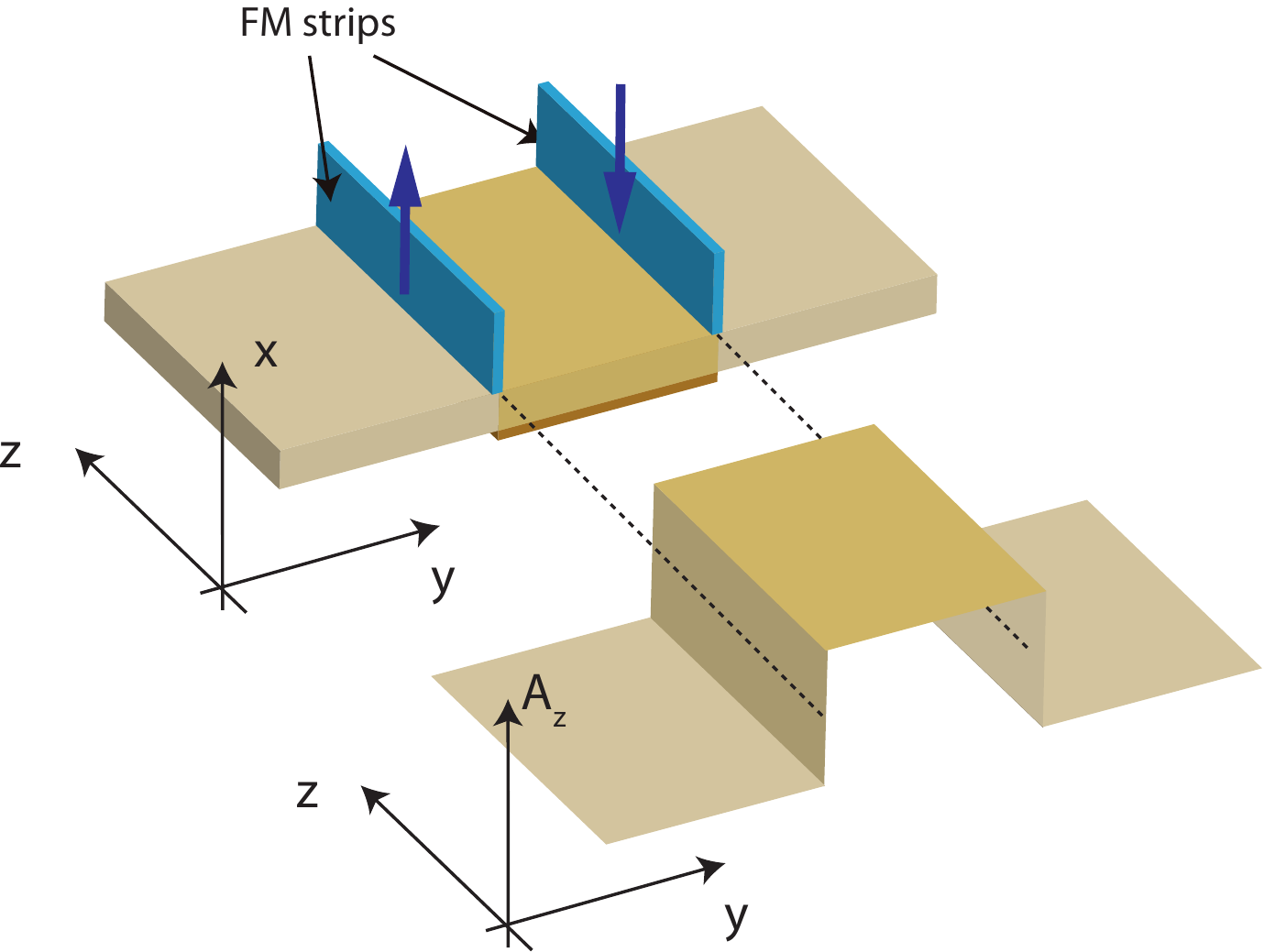}
\caption{ (a) A schematic of the tunneling setup with the inclusion of antisymmetrically magnetized FM strips, and (b) a plot of the corresponding $z$ component of the EM vector potential which is proportional to the $k_z$ shift. } 
\label{gFMstrip}
\end{figure}

The influence of the overlap between the momentum ranges perpendicular to the interface spanned by the lead and central segment EECs on the transmission profile suggests that the transmission profile can be manipulated by changing the overlap range. One way of achieving this is to introduce a shift of the EEC profile in momentum space by introducing antisymmetrically magnetized ferromagnetic (FM) strips to the interfaces between the leads and the central segment as shown in Fig. \ref{gFMstrip}.  The FM strips produce localized magnetic fields which we model as Dirac delta functions. The vector potential in the central segment between the strips then takes the form of a step function. The effects of the magnetic field can be incorporated via the replacement of $k_i \rightarrow \pi_i \equiv (k_i - A_i) = (k_i + \delta k_i)$ where $A_i$ is the electromagnetic vector potential. We choose the gauge such that the electromagnetic vector potential only has components along the $k$ direction we want to displace. The FM strips hence lead to a translation of the EEC of the central segments in $k$ space perpendicular to the direction of the FM magnetization. 

\begin{figure}[ht!]
\centering
\includegraphics[scale=0.6]{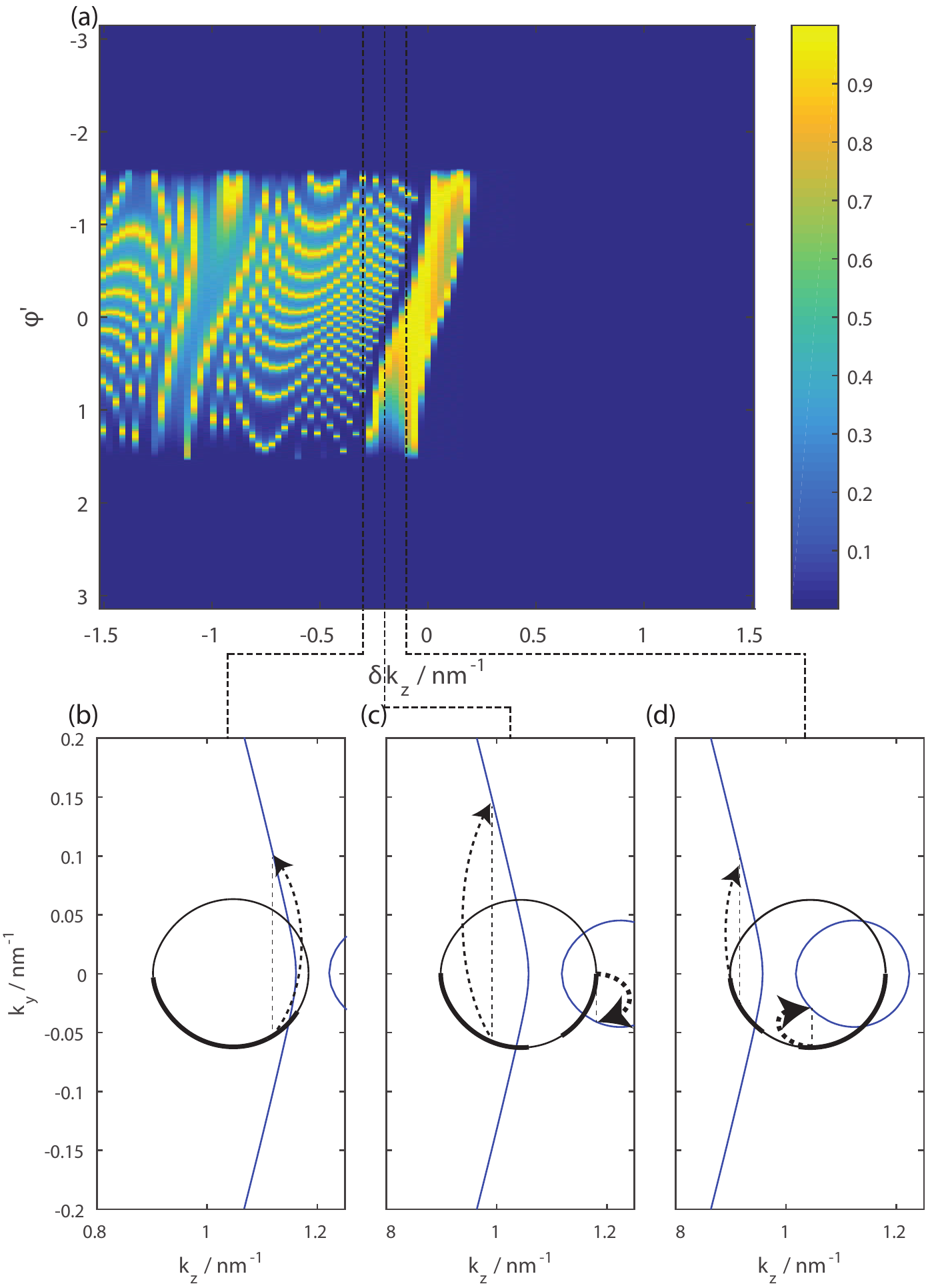}
\caption{ (a) The transmission as a function of the source angular coordinate $\varphi'$, and the central segment $k_z$ shift $\delta k_z$ at $E_2=60\ \mathrm{meV}$.  (b) to (d) show the source (black) and central segment (blue) EECs at the $\delta k_z$ values indicated on the plot. The states on the source EEC propagating in the $+y$ direction with ranges of $k_z$ that overlap with the $k_z$ range of states on the central segment propagating in the $+y$ direction are indicated as thicker lines. Note the narrow $\varphi'$ ranges in (c) and (d) where part of the source EEC falls into the gap between the central segment bulk and Fermi arc states. The thickness of the arrows pointing from the source to the central segment EECs in (b) to (d)  show the relative probability of transmission form the source states to the central segment states at the value of $k_z$ indicated by the dotted lines.  } 
\label{gCA15comb}
\end{figure}

Fig. \ref{gCA15comb} shows the transmission from such a setup with fixed $E_2=60\ \mathrm{meV}$ and length $100\ \mathrm{nm}$ as a function of the FM strip induced $k_y$ shift $\delta k_y$ and angular coordinate $\varphi'$ around the source EEC. 
 
The transmission is zero for large positive values of $\delta k_z$ when the largest value of $k_z$ for the drain propagating states at $k_y=0$ on the bulk Fermi arcs are pushed beyond the smallest value of $k_z$ on the source EEC, resulting in 0 overlap between the source and drain EECs. The transmission profile for smaller, and negative values of $\delta k_z$ in the figure can be divided into three regime -- a narrow band of $\delta k_z$ range where the transmission is near unity and oscillates weakly, an even narrower band of 0 transmission as $k_z$ decreases further, and a broad band of $k_z$ range where the transmission shows great oscillation with $\delta k_z$. 

The narrow band of $\delta k_z$ with near unity transmission corresponds to the $k_z$ range where $k_z$ values of the source and drain bulk EECs overlap. (They overlap partially in panels (c) and (d) of Fig. \ref{gCA15comb}. The transmission is then contributed largely by the transmission from source to central segment bulk states, which occurs at a higher probability than the transmission from source bulk states to central segment Fermi arc states. Within this energy band the transmission peaks for a given $\varphi$ when $\delta k_z$ compensates exactly for the $E_2$ shift in the central segment Dirac cone cross sectional ellipse so that the pseudospin orientation between the source and central segment EECs at a given value of $k_z$ are aligned.

As $\delta k_z$ decreases we encounter the situation (panel (c) and (d) ) where a narrow $\varphi$ range on the source EEC falls into the $k_z$ range falling in between the source bulk and Fermi arc states. There is 0 transmission in this $\varphi$ range since there are no propagating central segment states at the corresponding values of $k_z$.  

A further decrease in $\delta k_z$ leads to more, and eventually, all of the $k_z$ range spanned by the source EEC spanned into that spanned by the central segment Fermi arc EECs. The relatively large $k_y$ vector differences between the source and central segment states propagating in the $+z$ direction leads to the more pronounced transmission oscillations. 
 
\section{Conclusion}
In this work we studied the transmission of the hole states of the positive $k_z$ valley in  \ce{Na3Bi} thin film from a source lead through a central segment with a gate potential to an identical drain lead.  The finite thickness of the thin film results in the appearance of Fermi arc states as well as energy subbands in the dispersion relation. We considered the two cases where the interfaces between the leads and the central segment are (i) perpendicular and (ii) parallel to the $k$ space separation in the $k_z$ direction between the Weyl nodes. We saw that in the first case, inter-valley scattering gives rise to pronounced transmission oscillations with the variation of the central segment length and gate potential while the Fermi arc results in finite transmission even when the energy falls within the central segment bulk energy gap. Both of these features are not captured by earlier studies based on models which consider only a single Weyl node. We also saw that for interfaces parallel to the $z$ direction, the Fermi arc states result in a finite transmission even when there is zero transverse momentum overlap between the source and central segment \textit{bulk} states. This transmission can be modulated by the introduction of thin ferromagnetic strips at the central segment-lead interfaces where the electromagnetic vector potential leads to a translation of the central segment equal energy contours in $k$ space. 

\section{Acknowledgments} 
We thank the MOE Tier II grant MOE2013-T2-2-125 (NUS Grant No. R-263-000-B10-112), and the National Research Foundation of Singapore under the CRP Program ''Next Generation Spin Torque Memories: From Fundamental Physics to Applications'' NRF-CRP9-2013-01 for financial support.

\end{document}